\documentclass[11pt]{article}

\usepackage{mathptmx} 
\usepackage{graphicx}
\usepackage{lscape}
\usepackage{amsmath}
\usepackage{verbatim}
\usepackage{cite}
\usepackage{float}
\usepackage{color}

\usepackage[labelfont=bf,labelsep=period,justification=raggedright]{caption}

\makeatletter
\renewcommand{\@biblabel}[1]{\quad#1.}
\makeatother

\makeatletter
\newcommand{\customlabel}[2]{%
\protected@write \@auxout {}{\string \newlabel {#1}{{#2}{}}}}
\makeatother

\date{}

\begin{document}

\customlabel{fig:Ala_Rama_HSV_ALL_edited}{S1}
\customlabel{fig:Ala_Rama_HSV_GENEROUS_edited}{S2}
\customlabel{fig:Ala_Rama_HSV_CORE_edited}{S3}
\customlabel{fig:Alanine_RSA_distribution}{S4}
\customlabel{tab:max_SA_complete}{S1}
\customlabel{tab:max_exposed_RAMA}{S2}
\customlabel{tab:hydro_scales}{S3}
\customlabel{tab:bin_cutoffs}{S4}

\begin{flushleft}
{\Large\textbf{Maximum allowed solvent accessibilites of residues in proteins}}
\\
\mbox{}\\
Matthew Z. Tien$^1$,
Austin G. Meyer$^{2,3}$,
Dariya K. Sydykova$^{2}$,
Stephanie J. Spielman$^{2}$,
Claus O. Wilke$^{2,\ast}$
\\
\mbox{}\\
\textbf{1} Dept.\ of Biochemistry \& Molecular Biology, The University of Chicago, Chicago, IL 60637, USA
\\
\textbf{2} Section of Integrative Biology, Institute for Cellular and Molecular Biology, and Center for Computational Biology and Bioinformatics, The University of Texas at Austin, Austin, TX 78731, USA
\\
\textbf{3} School of Medicine, Texas Tech University Health Sciences Center, Lubbock, TX, 79430, USA
\\
$\ast$ Email: wilke@austin.utexas.edu
\end{flushleft}


\section*{Abstract}
The relative solvent accessibility (RSA) of a residue in a protein measures the extent of burial or exposure of that residue in the 3D structure. RSA is frequently used to describe a protein's biophysical or evolutionary properties. To calculate RSA, a residue's solvent accessibility (ASA) needs to be normalized by a suitable reference value for the given amino acid; several normalization scales have previously been proposed. However, these scales do not provide tight upper bounds on ASA values frequently observed in empirical crystal structures. Instead, they underestimate the largest allowed ASA values, by up to 20\%. As a result, many empirical crystal structures contain residues that seem to have RSA values in excess of one. Here, we derive a new normalization scale that does provide a tight upper bound on observed ASA values. We pursue two complementary strategies, one based on extensive analysis of empirical structures and one based on systematic enumeration of biophysically allowed tripeptides. Both approaches yield congruent results that consistently exceed published values. We conclude that previously published ASA normalization values were too small, primarily because the conformations that maximize ASA had not been correctly identified. As an application of our results, we show that empirically derived hydrophobicity scales are sensitive to accurate RSA calculation, and we derive new hydrophobicity scales that show increased correlation with experimentally measured scales.


\section*{Introduction}

Relative solvent accessibility (RSA) has emerged as a commonly used metric describing protein structure in computational molecular biology, with the particular application of identifying buried or exposed residues. It is defined as a residue's solvent accessibility (ASA) normalized by a suitable maximum value for that residue.  RSA was first introduced in the context of hydrophobicity scales derived by computational means from protein crystal structures \cite{Chothia1976,Rose1985,Miller1987,Moelbert2004,Shaytan2009}. More recently, RSA has been shown to correlate with protein evolutionary rates and has been incorporated as a parameter into models which determine these rates \cite{Goldmanetal1998,Bloometal2006,FranzosaXia2009,Zhouetal2009,FranzosaXia2012,Scherrer2012,MeyerWilke2012,ConantStadler2009}. As RSA straightforwardly characterizes the local environment of residues in protein structures, many studies have developed computational methods to predict RSA from protein primary and/or secondary structure \cite{RostSander1994,Pollastrietal2002,KimPark2004,Adamczaketal2004,NguyenRajapakse2005,Petersenetal2009,SinghAhmad2009}. Further applications of RSA include identification of surface, interior, and interface regions in proteins \cite{Levy2010}, protein-domain prediction \cite{Chengetal2006}, and prediction of deleterious mutations \cite{ChenZhou2005}.
 
To derive a residue's RSA from its surface area, an ASA normalization factor is needed for each amino acid. By convention, these normalization values have been derived by evaluating the surface area around a residue of interest X when placed between two glycines, to form a Gly-X-Gly tripeptide. Most commonly, the normalization values utilized are those previously calculated by either Rose \emph{et al.}~\cite{Rose1985} or Miller \emph{et al.}~\cite{Miller1987}. The primary distinction between these two sets of normalization values lies in the different $\phi$ and $\psi$ dihedral backbone angles chosen when evaluating Gly-X-Gly tripeptide conformations. Rose \emph{et al.}~\cite{Rose1985} considered tripeptides with backbone angles representing an average of observed $\phi$ and $\psi$ angles, whereas Miller \emph{et al.}~\cite{Miller1987} considered tripeptides in the extended conformation ($\phi= -120^\circ$, $\psi=140^\circ$).

As the number of empirically determined 3D protein crystal structures has grown over the years, it has become apparent that neither the Rose~\cite{Rose1985} nor the Miller~\cite{Miller1987} scale accurately identifies the true upper bound for a residue's ASA. In fact, virtually all amino acids display, on occasion, ASA values in excess of the normalization ASA values provided by either scale. Some do so quite frequently (e.g.\ R, D, G, K, P), reaching RSA values of up to 1.2. This discrepancy, which leads to RSA values $>1$, is generally known in the field though rarely acknowledged in print. One exception is a recent study that carried out an extensive empirical survey of ASA values in PDB structures  \cite{SinghAhmad2009}. That study found that the most accessible conformations are generally found in loops and turns, not in the extended conformation, and it suggested to use conformation-dependent maximum ASA values for normalization \cite{SinghAhmad2009}.

Here, we derive a new set of ASA normalization values that provide a tight upper bound on ASA values observed in biophysically realistic tripeptide conformations. To calculate these normalization values, we pursue two complementary strategies---one empirical and one theoretical. For the empirical approach, we mined thousands of 3D crystal structures and recorded the maximum ASA values we found for each amino acid across all structures. For the theoretical approach, we computationally built Gly-X-Gly tripeptides and systematically evaluated all biophysically allowed conformations to determine a maximum theoretical ASA value. These two strategies yield congruent results and ultimately produce comparable normalization scales that tightly bound ASA for all 20 amino acids. We then return to the historic motivation for RSA and investigate the implications of our results for hydrophobicity scales. We find that ASA normalization affects the performance of empirically derived hydrophobicity scales, and we propose new scales that show improved correlation with experimentally measured scales.

\section*{Results}

\subsection*{Published ASA normalization values are too small}
\begin{figure}[H]
\includegraphics[width=4in]{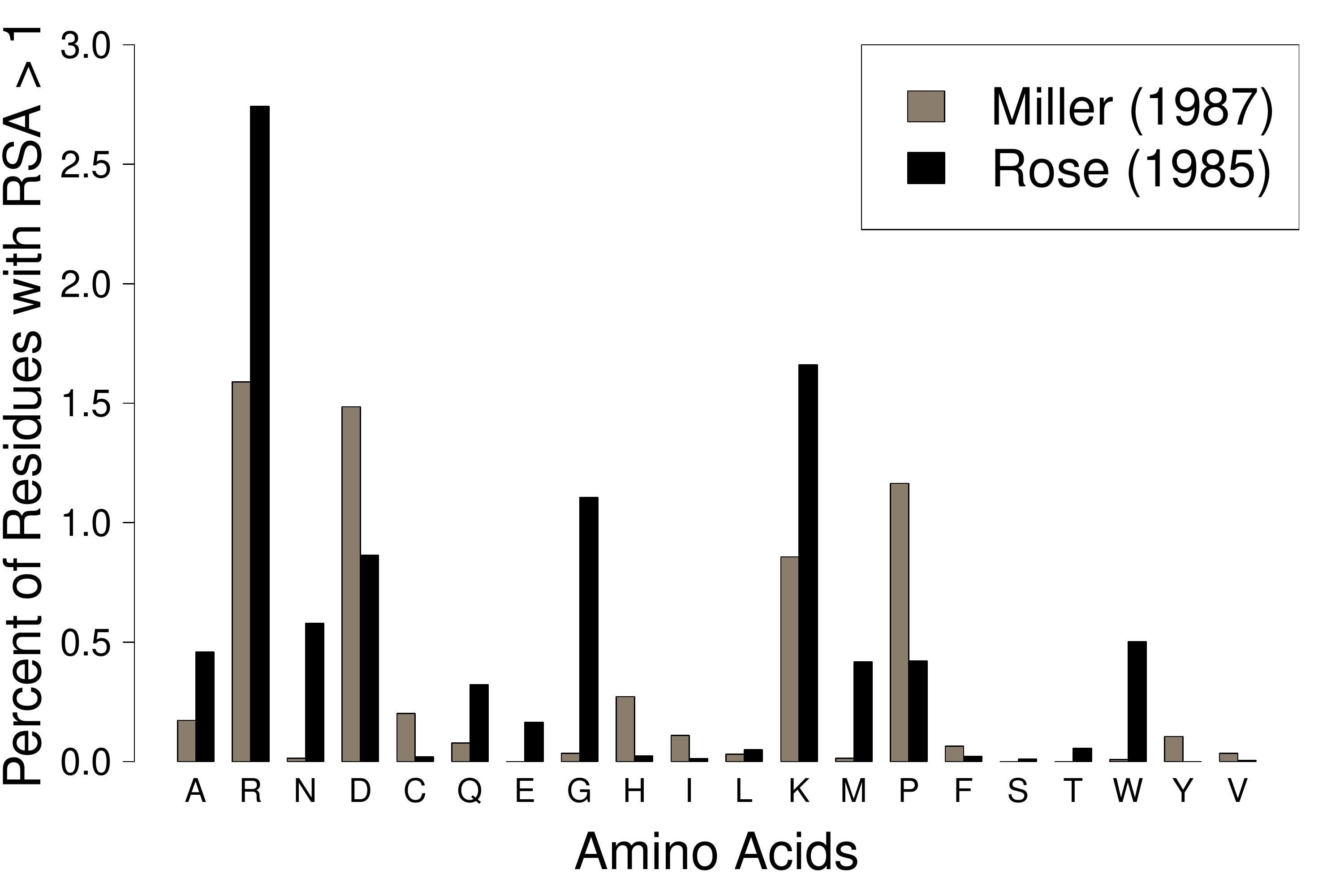}
\caption{\label{fig:BarGraphRSA}Frequency of residues with $\text{RSA}>1$ in empirical protein structures. Nearly all amino acids, and notably R, D, K, G, and P, show $\text{RSA}>1$ 
when RSA is calculated using the normalization values of either Rose \emph{et al.} \cite{Rose1985} or Miller \emph{et al.} \cite{Miller1987}.}
\end{figure}

We initially assessed the accuracy of Rose's~\cite{Rose1985} and Miller's~\cite{Miller1987} ASA normalization scales through an exhaustive survey of the ASA values found in experimentally determined protein structures. We obtained a list of 3197 high-quality PDB structures from the PISCES server \cite{WangDunbrack2003}. We then calculated ASA for each residue in all 3197 structures, excluding any chain-terminating residues. ASA values were subsequently normalized using the scales of either Rose \emph{et al.}~\cite{Rose1985} or Miller \emph{et al.}~\cite{Miller1987} to obtain RSA. For either scale and each amino acid, we found that residues with $\text{RSA}>1$ were not uncommon (Figure~\ref{fig:BarGraphRSA}); RSA values exceeded unity by up to 20\%. The amino acids that most commonly displayed $\text{RSA}>1$ were R, D, G, K, P. For those amino acids, RSA values $>$1 occurred at frequencies of 1\% to 3\% of all residues, depending on the normalization scale used (Figure~\ref{fig:BarGraphRSA}).

\begin{figure}[H]
\includegraphics[width=3in]{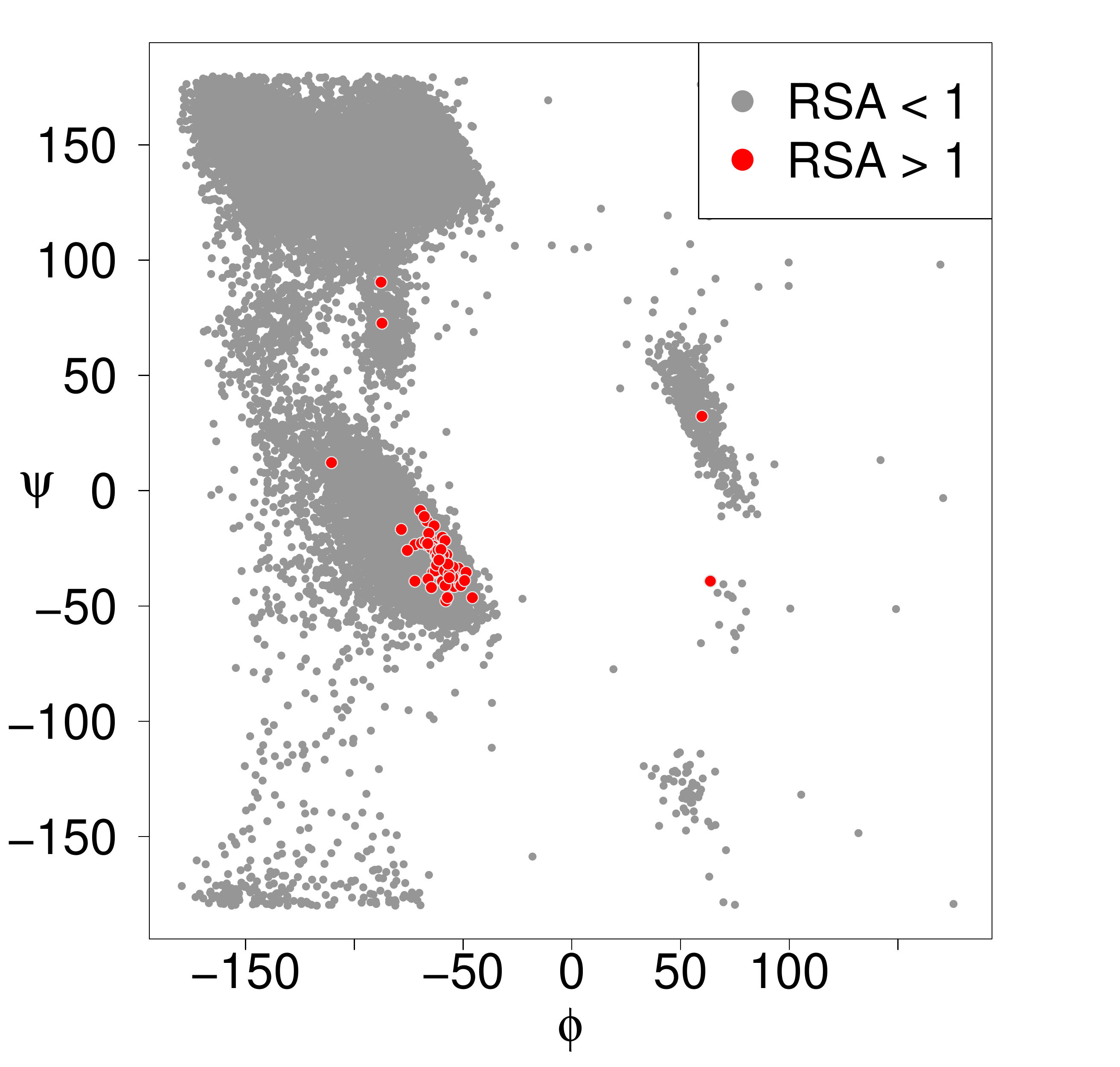}
\caption{\label{fig:RamaAla}Ramachandran plot for alanine residues in our empirical data set. Coordinates which correspond to RSA values $>1$ are shown in red and are clearly concentrated around coordinates $(-50^\circ,-45^\circ)$. We therefore propose that this region contains the maximally exposed conformation of alanine and should be used for calculating maximum ASA.}
\end{figure}

To determine the underlying factors leading to $\text{RSA}>1$, we examined the association between RSA and the following factors: residue neighbors, secondary structure, bond lengths, bond angles, and dihedral angles. For most of these quantities, we found no strong association with RSA. We did, however, find a clear association with residues' $\phi$ and $\psi$ backbone angles. For example, consider the Ramachandran plot of alanine (Figure~\ref{fig:RamaAla}). A noticeable cluster of high-RSA residues falls into the $\alpha$-helix region of $\phi\approx-50$, $\psi\approx-45$. We found similar results for all other amino acids. Importantly, neither Rose nor Miller derived their normalization ASA values in that region of backbone angles. Therefore, we concluded that previous ASA normalization scales were obtained with poorly chosen $\phi$ and $\psi$ angles.

\subsection*{Modeling Tripeptides Yields Significantly Higher Maximum ASA Values}
To derive maximum ASA values for each amino acid X, we computationally constructed Gly-X-Gly tripeptides and systematically rotated them through all biophysically allowed  conformations (see Methods and Supporting Text for details.) When constructing the tripeptides, we set bond lengths and angles (excluding $\omega$, $\phi$, $\psi$, and $\chi$ angles) for each amino acid equal to the average values observed for that amino acid in our reference set of 3197 PDB structures. We set $\omega=180^\circ$. We then rotated the $\phi$ and $\psi$ around the X residue in  discrete 1$^\circ$ steps, exhaustively enumerating all conformations. Additionally, we iterated through all rotamer angles $\chi$ that were sterically possible with each $(\phi, \psi)$ combination. For those amino acids with more than 10 possible distinct rotamer conformations, as determined by the Dunbruck database \cite{WangDunbrack2003}, we evaluated ten randomly chosen rotamer conformations. We recorded the maximum ASA observed for each $(\phi, \psi)$ backbone-angle combination.

\begin{figure}[H]
\includegraphics[width=6in]{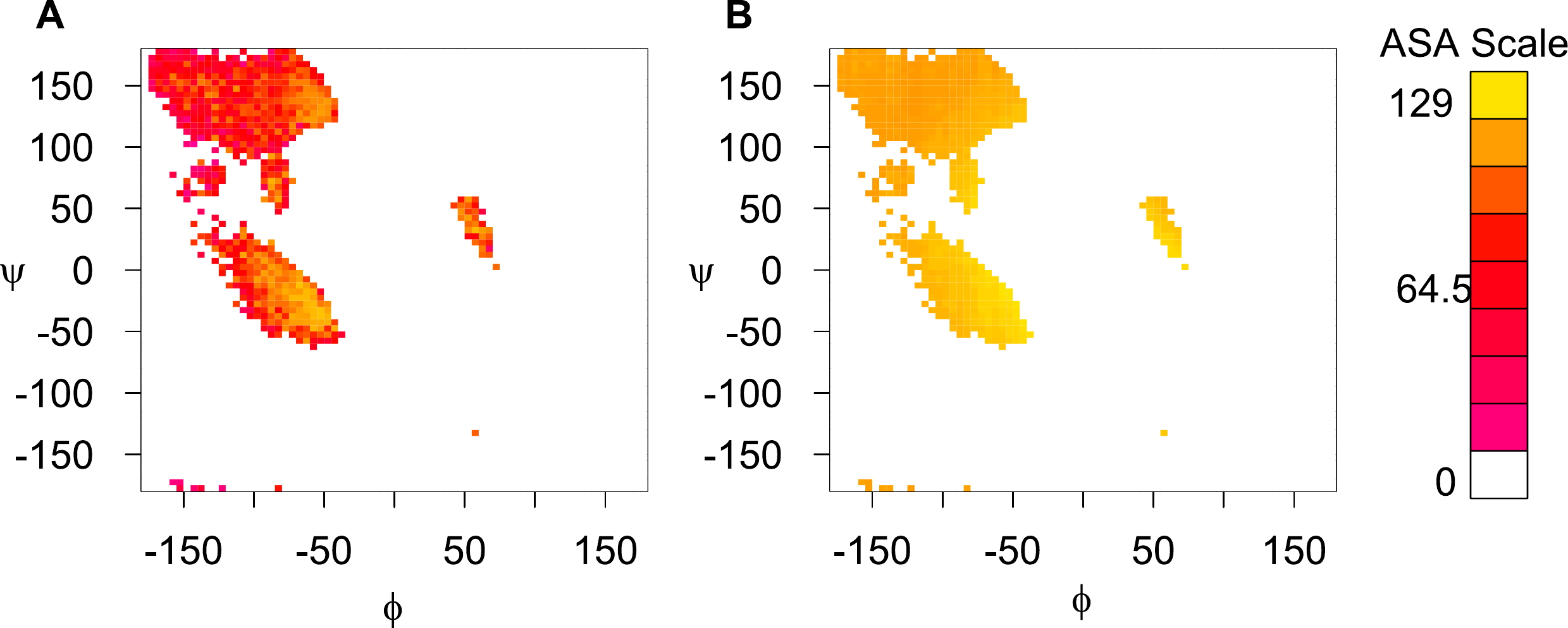}
\caption{\label{fig:heatrama} Ramachandran plots for empirical and theoretical maximum ASA values of alanine. (A) Empirical maximum ASA values for each $5^\circ$ by $5^\circ$ bin. All bins in the ALLOWED region are shown. (B) Theoretical maximum ASA values, as determined by computational modeling, shown for non-empty bins in (A). Both the empirical and the theoretical approach find the largest ASA values in the $\alpha$-helix region around $(-50^\circ,-45^\circ)$. By contrast, the extended conformation $(-120^\circ, 140^\circ)$ leads to relatively low maximum ASA.}
\end{figure}

Next, we compared the resulting theoretical maximum ASA values to the empirically observed maximum ASA values. We binned both the theoretical and the empirical values into discrete $5^\circ\times5^\circ$ bins of $(\phi, \psi)$ and recorded the maximum ASA in each bin. To eliminate nonexistent or rare conformations, we defined four Ramachandran regions for each amino acid: CORE, containing at least 80\% of the empirical observations; ALLOWED, containing at least 97\% of the empirical observations; GENEROUS, extending the core region by $20^\circ$ in all directions; and ALL, containing all non-empty bins. The definitions of the CORE, ALLOWED, and GENEROUS regions are consistent with the definitions used in Ref.~\cite{Morrisetal1992}. For each region, we displayed the maximum ASA value in each bin in side-by-side Ramachandran plots (Figures~\ref{fig:heatrama} and \ref{fig:Ala_Rama_HSV_ALL_edited}-\ref{fig:Ala_Rama_HSV_CORE_edited}) and generally found good congruence between the theoretical and the empirical values for all amino acids. Regions that had the highest maximum ASA in the theoretical data set also had the highest maximum ASA in the empirical data set. The highest ASA values were generally observed in the $\alpha$-helix region of the Ramachandran plot (Figure~\ref{fig:heatrama}). Based on these results, we propose new maximum ASA values (Tables~\ref{tab:results} and \ref{tab:max_SA_complete}) and maximally exposed geometries for each amino acid (Table~\ref{tab:max_exposed_RAMA}).

\begin{figure}[H]
\includegraphics[width=3in]{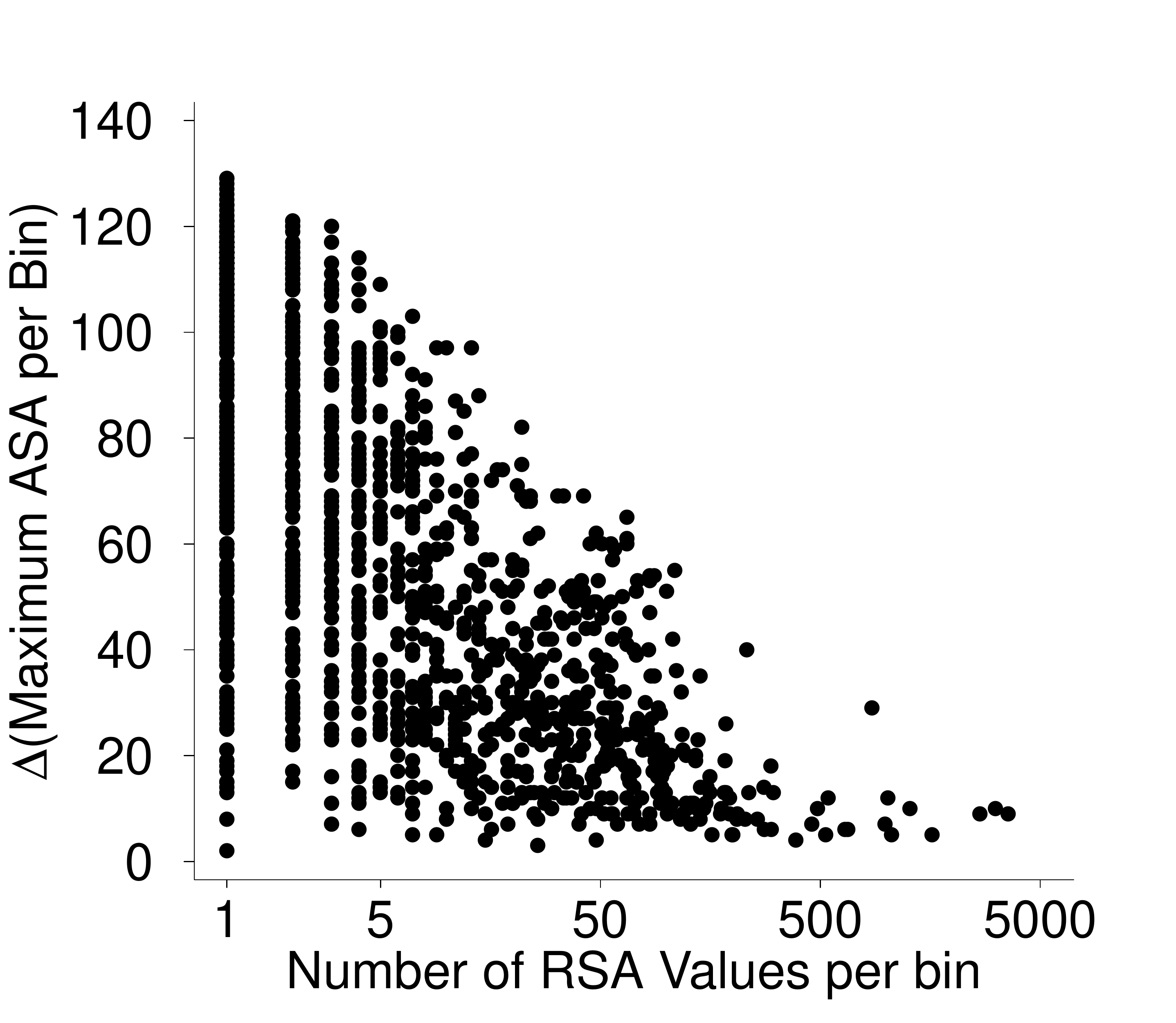}
\caption{\label{fig:EvC} Difference between theoretically and empirically determined maximum ASA values for alanine, across $5^\circ$ by $5^\circ$ bins. As the amount of data per bin increases, the difference between theoretical and empirical maximum ASA approaches zero, demonstrating that our two methods converged with increasing amounts of data. Furthermore, the difference between values is frequently close to zero, even when little data is available for a bin. This observation indicates that our theoretically derived maximum ASA values provide a tight bound on the empirically observed ones.}
\end{figure}

We further evaluated our model's performance by directly comparing theoretical and empirical maximum ASA values in each $(\phi, \psi)$ bin. We calculated the difference between these two values for each $5^\circ\times5^\circ$ bin (now including all bins with at least one observation in the empirical data set). We then plotted this difference against the number of empirical observations obtained for each bin (Figure~\ref{fig:EvC}). We found that with increasing amounts of empirical data, this difference approached zero; the maximum ASA values from both approaches converged as more data was available. Moreover, even for sparsely populated bins, at least some bins showed a difference near zero, regardless of the number of observations in each bin. Therefore, while our results did improve with increasing amounts of data, they were also largely robust to smaller data sets.

As Table~\ref{tab:max_SA_complete} shows, the maximum ASA values observed in the empirical data set were nearly identical for different Ramachandran regions. Scales for the ALLOWED, GENEROUS, and ALL regions were identical, with the exception of a 1 \AA$^2$ difference for Val between ALLOWED and GENEROUS/ALL. The scale for the CORE region was nearly identical as well, with most differences on the order of 1-2 \AA$^2$. The only larger difference (15 \AA$^2$) arose for Cys, the rarest amino acid in our data set. For the theoretical scales, we similarly found that differences between the CORE and ALLOWED regions were minor, typically on the order of 2-5 \AA$^2$. The biggest difference again arose for Cys. Theoretical maximum values in the GENEROUS and ALL regions were up to 10-15 \AA$^2$ larger than in the ALLOWED region, and generally substantially larger than the largest ASA values observed in the entire empirical data set. We conclude from this finding that the GENEROUS and ALL regions are too permissive of unphysical and/or rare backbone conformations, and we recommend that the maximum ASA values of the ALLOWED region be used in actual applications. Table~\ref{tab:results} summarizes these values and compares them to the previously published scales by Miller \emph{et al.}~\cite{Miller1987} and Rose \emph{et al.}~\cite{Rose1985}. All results in the remainder of this work were derived using the scales obtained for the ALLOWED region.

\subsection*{Relation to Empirically Derived Hydrophobicity Scales}

The solvent exposure of an amino acid, averaged over many occurrences of that amino acid in many different protein structures, should correlate with the amino acid's hydrophobicity. Therefore, solvent exposure has long been used as a means to empirically derive hydrophobicity scales from protein crystal structures \cite{Chothia1976,Rose1985}. In particular, Rose \emph{et al.}~\cite{Rose1985} derived a hydrophobicity scale by calculating the mean RSA for each amino acid across a set of reference crystal structures, using the ASA normalization values derived in the same work~\cite{Rose1985}. Since those normalization values are inaccurate, as shown above, we assessed how using our normalization values would alter the Rose hydrophobicity scale.

We first compared the Rose scale to a number of experimentally derived scales (Table~\ref{tab:Corr}, \cite{Wolfenden1981,Kyte1981,Fauchere1983,Radzicka1988,MacCallum2007,Moon2011,Wimley1996}). We included in the list of experimental scales the scale by Kyte \& Doolittle~\cite{Kyte1981}, which is a hybrid scale partially based on solvent-accessibility data from protein structures, and the scale by Mac Callum \emph{et al.}~\cite{MacCallum2007}, which is based on molecular-dynamics simulations. A brief description of each scale is given in the legend to Table~\ref{tab:Corr}. The Rose scale correlated reasonably well (50\%-70\% of variance explained) with most experimental scales. It correlated the highest with the scale of Fauchere \& Pliska~\cite{Fauchere1983} (82\% of variance explained) and it did not correlate significantly with the scales of Wimley \emph{et al.}~\cite{Wimley1996} and of Mac Callum \emph{et al.}~\cite{MacCallum2007} (Table~\ref{tab:Corr}).

We next derived two scales based on mean RSA, calculated using either our theoretical or our empirical ASA normalization values (Table~\ref{tab:hydro_scales}). Both of our mean RSA scales correlated well with the Rose scale ($r=0.96$ and $r=0.97$, respectively, with $P<10^{-10}$ in both cases) but were not identical to it. The biggest difference arose for histidine, which is ranked as the 8th-most hydrophobic amino acid according to the Rose scale but as the 10th- or 13th-most hydrophobic amino acid, respectively, according to our scales. Our scales correlated more strongly than the Rose scale with all experimental scales except the Mac Callum scale, which did not correlate significantly with either our or the Rose scale (Table~\ref{tab:Corr}). For the majority of experimental scales, the percent variance explained increased by approximately 10 percentage points using our normalization over the Rose normalization. We can conclude from these results that mean RSA is a useful measure of amino acid hydrophobicity and that correct ASA normalization is required to assign appropriate hydrophobicity scores to all amino acids.

One concern with using mean RSA as a measure of hydrophobicity is that the RSA distribution of individual amino acids tends to be highly skewed (see Figure~\ref{fig:Alanine_RSA_distribution} for an example). Hence, mean RSA may not accurately reflect the most common RSA values. It might be preferable to use instead the fraction of times an amino acid occurs in a buried conformation in empirical protein structures. This approach was originally suggested by Chothia \emph{et al.} in 1976 and executed with the limited data available at the time~\cite{Chothia1976}.

We calculated two additional scales from our data set of 3197 protein structures: for each of the 20 amino acids, we calculated the fraction of completely buried residues (100\% buried, $\text{RSA}=0$) and the fraction of 95\% buried residues ($\text{RSA}<0.05$) among all occurrences of these amino acids in the protein structures. For most of the experimental scales, these two scales showed a stronger correlation than any of the scales based on mean RSA did (Table~\ref{tab:Corr}). The two main exceptions were the scale by Fauchere \& Pliska \cite{Fauchere1983}, which correlated better with mean RSA, and the scales by Wimley \emph{et al.} \cite{Wimley1996} and by Mac Callum \emph{et al.}~\cite{MacCallum2007}, which correlated poorly with all empirical scales. Since the Kyte \& Doolittle scale \cite{Kyte1981} is partly based on the fraction of buried residues, its strong correlation with our scales is not surprising and does not represent a truly independent validation of these scales.

\section*{Discussion}

We have derived significantly improved ASA normalization values. Our normalization values provide a tight upper bound to the largest observed ASA values in empirical structures. By contrast, previously published ASA normalization valules were too small, by up to 20\%, and frequently led to RSA values $>1$. We estimated the maximum allowed ASA for each amino acid by computationally modeling Gly-X-Gly tripeptides, where X is the amino acid of interest, and exhaustively surveying ASA over all biophysically feasible conformations. We found that maximally exposed conformations tend to fall into the $\alpha$-helix region of Ramachandran plots, and that extended conformations display some side-chain burial. The results of our modeling approach were consistent with maximum ASA values found by surveying over 3000 empirical protein crystal structures. We also revisited the problem of deriving empirical hydrophobicity scales from protein structures. We found that improved ASA normalization values lead to improved empirical hydrophobicity scales. Further, scales based on both mean RSA and on the fraction of buried residues correlated well with experimentally measured scales. Overall, the fraction of 95\% buried residues seems to be the best-performing empirical hydrophobicity scale, but mean RSA correlates well with an experimental scale based on side-chain transfer between octanol and water.

Our method of obtaining ASA normalization values was similar to the methods employed by Rose \emph{et al.}~\cite{Rose1985} and by Miller \emph{et al.}~\cite{Miller1987}. Rose \emph{et al.}~\cite{Rose1985} calculated their ASA normalization values by computing the ASA of residue X in Gly-X-Gly tripeptides whose conformations were chosen based on the average dihedral angles from available empirical data at the time. Miller \emph{et al.}~\cite{Miller1987}, on the other hand, calculated their ASA normalization values by computing the ASA of an extended trimer structure with $\phi = -120^\circ, \psi= 140^\circ$ and with side-chain conformations that were frequently observed in the empirical data. The key distinction between these previous approaches and ours lies in our exhaustive sampling of tripeptide conformations. By modeling all biophysically feasible discrete combinations of $\phi$ and $\psi$ angles and varying rotamers, we identified the ideal conformations which yield maximum allowed ASA. To pursue our modeling strategy, we developed a program that allowed us to easily construct peptide chains from scratch in arbitrary conformations (see Supporting Text for details).

Our results are broadly consistent with a recent paper by Singh and Ahmad~\cite{SinghAhmad2009}. These authors did an extensive empirical survey of ASA values in tripeptides from PDB structures. They found that the highest observed ASA values were found in loops and turns, not in the extended conformation used by Miller \emph{et al.}. Their highest ASA values are generally consistent with ours. Further, Singh and Ahmad found that the highest observed ASA values were dependent on the neighboring residues around the focal residue. Finally, Singh and Ahmad showed that for RSA prediction from primary sequence, prediction accuracy could be improved by approximately 10\% if ASA values were normalized by (neighbor-dependent) highest observed ASA values rather than by ASA values observed in the extended conformation~\cite{SinghAhmad2009}. Our work serves as a useful complement to their work, by (i) providing, through molecular modeling, highest \emph{possible} ASA values rather than just highest \emph{observed} ASA values, by (ii) providing highest observed and highest possible ASA values as a function of backbone dihedral angle, and by (iii) demonstrating that improved RSA normalization yields empirical hydrophobicity scales that are more similar to experimentally measured ones.

In our modeling approach, we calculated ASA values for Gly-X-Gly tripeptides. Other authors have considered normalizations based on Ala-X-Ala tripeptides \cite{Ahmadetal2003,NguyenRajapakse2005} or even neighbor-specific normalizations (i.e., a different normalization for each specific tripeptide \cite{SinghAhmad2009}). We chose Gly-X-Gly tripeptides because we wanted to calculate the highest possible ASA values of tripeptides, and glycines will generally occlude less solvent than alanines. From a practical perspective, we prefer a simple normalization scheme, and hence highest possible ASA values are attractive to us. However, for certain applications, it may be the case that neighbor-specific or backbone-specific normalizations are preferable. Singh and Ahmad \cite{SinghAhmad2009} provided neighbor-specific normalization values, but didn't control for backbone angles. We have shown here that maximum ASA values depend substantially on backbone angles (e.g.\ Fig.~\ref{fig:heatrama}), and we provide both highest observed and highest possible ASA values as a function of backbone angles (see ``Data and code availability'' in Methods). It is not known at this time whether neighbor-dependent or backbone-dependent normalization is preferable, and the answer may depend on the specific application. In principle, one could also normalize by both neighboring amino acids and backbone dihedral angles. A modeling approach such as ours could be employed to calculate the highest possible ASA values for any tripeptide in any conformation. The computational resources required would be substantial, however, since we would have to model 400 times more tripeptides than we did for the present work.

Our theoretical modeling approach to exhaustively survey tripeptides has two potential shortcomings. First, for bond lengths and angles (except major dihedral angles), we used mean values observed in a large number of protein crystal structures. This approach neglects the variation around the mean, and there could be rare cases where unusually large bond lengths or unusual bond angles might cause ASA to become larger than estimated here. Such scenarios would have to be exceedingly rare, however, since we did not find a single case in which the largest empirically derived maximum ASA value exceeded the largest theoretically derived maximum ASA value (Table~\ref{tab:results}). Second, for amino acids with more than 10 distinct rotamer conformations, we did not exhaustively enumerate all possible conformations but only sampled 10 conformations at random. Thus, in principle it is possible that we missed a particular rotamer conformation that would have corresponded to a larger ASA value than the maximum we observed. Two arguments suggest that this issue is not likely a major source of error. First, again, we did not find a single case in which the empirical maximum ASA was larger than the theoretical maximum ASA. Second, maximum ASA varied slowly with $\phi$ and $\psi$, and by exhaustively enumerating conformations in $1^\circ$ steps, in effect we sampled the most exposed conformations multiple times, thus reducing the chance of missing a rare, large-ASA conformation.

As our RSA calculations are based on ASAs of tripeptides, we excluded all chain terminating residues from both the empirical and the theoretical analysis. Even with our improved ASA normalization values, then, chain-terminating residues may still display $\text{RSA}>1$. We therefore recommend that future analyses making use of RSA similarly exclude any chain-terminating residues, as their RSA estimates will not be precise. Suitable normalization values for chain-terminating residues are not available at present. 

The normalization values we have derived here are, strictly speaking, only valid for solvent-accessible surface areas calculated with the DSSP program \cite{Kabsch1983}. However, more generally, we expect them to be correct as long as solvent accessibility is calculated according to the definition of Lee and Richards \cite{LeeRichards1971}, which assumes that a sphere of radius 1.4\AA\ is rolled over the surface of the molecule. For cases in which solvent accessibility is calculated differently, our results suggest that one can follow an empirical approach to normalization. In other words, one need not exhaustively evaluate tripeptides, as we have done here. Instead, one can obtain a representative sample of structures from the protein data bank, exclude all terminal residues and residues in unusual conformations, and then find for each amino acid the maximum solvent accessibility within that data set, according to one's chosen definition of solvent accessibility. As Table~\ref{tab:results} shows, this empirical approach should generally yield results that are quite similar to the theoretical normalization values.

In many applications, specifically in the context of sequence evolution, RSA is treated as a site-specific property that is invariant under mutation. While RSA values of homologous structures tend to be strongly correlated \cite{RostSander1994,Zhouetal2009}, individual sites, in particular exposed ones, can show substantial RSA variability \cite{RostSander1994}. In this context, we would like to emphasize that one potential source of RSA variability in previous studies was RSA normalization. For example, Ref.~\cite{RostSander1994} used the Rose scale, which differs quite substantially from the scale we propose here. In particular, the corrections we propose to the Rose scale range from 4\% (for Leu) to 18\% (for Asp), and are approximately uniformly distributed in that range over the 20 amino acids. Thus, one can envision scenarios under which a substitution that might not change RSA under our scale might change it by over 10\% under the Rose scale.
At the same time, we have to realize that RSA can show variability even in the absence of mutation, in particular for exposed residues. A residue in a surface loop will undergo thermodynamic fluctuations, and its solvent exposure state will vary over time as neighboring residues move closer in or further out. By contrast, a residue in the core will likely remain solvent-occluded at all times. To obtain a reliable RSA value for a surface residue, one would thus ideally calculate an average over a thermodynamic ensemble of structures. A detailed analysis of RSA variability under thermodynamic fluctuations and among homologous structures is beyond the scope of this work but should be undertaken in the future.

The comparison between experimentally and empirically derived hydrophobicity scales has been a persistent topic in biochemistry. As of this writing, the AAIndex database \cite{KawashimaKanehisa2000} contains over 40 scales related to amino acid hydrophobicity or polarity. While these scales tend to cluster \cite{TomiiKanehisa1996,Kawashimaetal2008}, there are substantial dissimilarities among hydrophobicity scales, and any two scales within the hydrophobicity cluster may not correlate that well. Any further insight into the mechanisms that cause differences among scales derived under different conditions or using different methodologies would improve our understanding of protein biochemistry. In particular, resolving discrepancies between empirically-derived data and experimentally derived thermodynamics of hydrophobicity could provide crucial insight into algorithms of protein-structure prediction and de-novo protein folding. 

Wolfenden \emph{et al.}~\cite{Wolfenden1981} were the first to propose an approach for reconciling the empirical and the experimental approach, by correlating the distribution of amino acid exposure with their experimental behaviors in water/vapor solutions. More recently, Moelbert \emph{et al.}~\cite{Moelbert2004} attempted to reconcile these disparities by correlating hydrophobic states with surface-exposure patterns of protein structures. Additionally, Shaytan \emph{et al.}~\cite{Shaytan2009} assessed the distribution of amino acid exposure in proteins to discern apparent free energies of transfer between protein interior and surface states, and found that free energy is highly correlated with experimental hydrophobicity scales \cite{Shaytan2009}. Each of these approaches used the ASA normalization values from either Rose \emph{et al.}~\cite{Rose1985} or Miller \emph{et al.}~\cite{Miller1987}. Since the normalization ASA values developed here are more accurate, we believe that our findings are valuable for determining exposure states. Using the Rose hydrophobicity scale as an example, we have shown here that improved ASA normalization values consistently yield improved correlations with experimental scales, irrespective of the exact type of experimental scale considered.  Of all empirical scales we analyzed, however, the fraction of 95\% buried residues was most consistently strongly correlated with different experimental scales and thus could be considered the overall best-performing empirical scale. 

Further, in agreement with Shaytan \emph{et al.}~\cite{Shaytan2009}, we found that different experimental scales corresponded to different empirical scales. For example, transfer energies from water to vapor correlated the strongest with the fraction of 100\% buried residues, while transfer energies from water to cyclohexane correlated the strongest with the fraction of 95\% buried residues, and transfer energies from water to octanol correlated the strongest with mean RSA. Since mean RSA puts more weight on exposed residues than does the fraction of either 100\% buried or 95\% buried residues, this finding agrees with the three distinct types of scales found by Shaytan \emph{et al.}~\cite{Shaytan2009}. The pentapetide scale by Wimley \emph{et al.}~\cite{Wimley1996}, however, did not correlate well with either of the empirical scales we considered. Wimley \emph{et al.} performed a partioning experiment between water and 1-octanol using pentapeptide species, Ace-WLXLL, with X being one of the naturally occuring 20 amino acids. Otherwise, their set up was simlar to the one of Fauchere \& Pliska~\cite{Fauchere1983}. By using pentapeptides rather than individual amino acids, the Wimley \emph{et al.} hydrophobicity scale does not seem to accurately reflect the hydrophobic character of individual amino acids but rather that of the pentapeptides.

In summary, we have presented significantly improved ASA normalization values. We recommend that our theoretical normalization values for the ALLOWED region (column 1 of Table~\ref{tab:results}) be used to normalize ASA. The optimal hydrophobicity scale will depend on the specific application, but the fraction of 95\% buried residues seems to be the best general-purpose empirical scale.

\section*{Materials and Methods}

\subsection*{Empirical maximum ASA values}

We obtained a set of 3197 high-quality protein crystal structures using the PISCES server \cite{WangDunbrack2003}. We imposed the following requirements: resolution of 1.8 \AA\ or less, an $R$-free value $<0.25$, and a pairwise mutual sequence identity of at most 20\%. For each amino-acid residue in all 3197 structures, we retrieved bond lengths, bond angles, dihedral angles, peptide bond lengths, and nearest neighbors. Chain-terminating residues, defined as those residues whose peptide bond lengths with any neighboring residue was greater than six standard deviations from the protein's mean peptide bond length, were excluded from all subsequent analyses. We further identified all residues in the data set that had either missing atoms or atoms with ambiguous occupancy data (PDB occupancy column contained a number $<1.0$ for at least one atom in the residue). We eliminated these residues and their immediate neighbors from all subsequent analyses as well.

We used the program DSSP (2011 version) \cite{Kabsch1983} to calculate solvent accessibility (ASA) and to identify the secondary structure of each residue across all proteins. Because of the quality control we imposed on residues (see preceding paragraph), our final ASA data set only contained residues that were complete and unambiguous and whose neighbors were complete and unambiguous as well.

We next filtered by allowed Ramachandran angles. For each amino acid, we binned all observed $\phi, \psi$ combinations into $5^\circ\times5^\circ$ squares, and assigned each square to one or more of the following regions: The CORE region was defined to contain at least 80\% of the observed Ramachandran angles. The ALLOWED region was defined to contain at least 97\% of the observed Ramachandran angles. For both the CORE and the ALLOWED regions, we identified, for each amino acid, the number of observations per $5^\circ\times5^\circ$ bin required for that bin to be part of the respective region. Table~\ref{tab:bin_cutoffs} lists these bin cutoffs. The GENEROUS region was defined to extend the ALLOWED region by $20^\circ$ in all directions, regardless of whether the particular Ramachandran angles have been observed. Finally, the ALL region was defined to contain all observed Ramachandran angles. The definitions of the CORE, ALLOWED, and GENEROUS regions are consistent with current biochemical convention \cite{Morrisetal1992,Laskowskietal1993}. For all four regions, we identified the maximum ASA observed.

We calculated RSA as $\text{RSA}=\text{ASA}/\text{Maximum ASA}$, where ``Maximum ASA'' corresponds to the maximum ASA value, as determined by the normalization scale used, for the focal amino acid.

\subsection*{Theoretical maximum ASA values}

To find the theoretical maximum solvent accessibility (ASA) for each amino acid X, we computationally constructed Gly-X-Gly tripeptides. Each tripeptide was modeled by specifying coordinates of each constituent atom, using bond lengths and angles from our empirically mined protein structures. Briefly, we first constructed peptides in a defined conformation by placing each atom at the correct position in 3D space. We then adjusted $\phi$, $\psi$, and $\chi$ angles to obtain the desired conformation. This method is described in more detail in Supporting Text, and the computer code to carry out tripeptide construction has been published as a stand-alone library \cite{Tienetal2013}. 

Once constructed, we exhaustively rotated $\phi$ and $\psi$ dihedral backbone angles in discrete $1^\circ$ increments, holding $\omega$ constant at $180^\circ$. For each $(\phi, \psi)$ combination, we additionally rotated through all possible $\chi$ rotamer angles, as found in the Dunbruck Rotamer Database \cite{WangDunbrack2003}. Rotamer angles were grouped into three $120^\circ$ sectors (60$^\circ$, -60$^\circ$, and 180$^\circ$) and averaged within each sector. For amino acids where the side chain could assume more than ten distinct rotamer conformations (e.g. for L, I, M, K, N), we selected ten rotamer conformations at random instead of exhaustively enumerating all rotamer conformations. A different set of randomly chosen rotamer conformations was generated for each combination of $(\phi, \psi)$ angles. 

For each tripeptide conformation examined, a corresponding PDB file was created and inputted into the program DSSP \cite{Kabsch1983} to compute the ASA of amino acid X. For each amino acid and $(\phi, \psi)$ combination, we recorded the largest ASA value from all rotamer variations examined. To determine the theoretical maximum ASA value for each amino acid, we identified the largest ASA value observed for any $(\phi, \psi)$ combination within one of the four Ramachandran regions defined above (CORE, ALLOWED, GENEROUS, ALL).

\subsection*{Hydrophobicity scales}

We calculated empirical hydrophobicity scales on the same set of 3197 crystal structures. Mean RSA of each amino acid was calculated as the RSA averaged over all occurrences of that amino acid in the data set. The corresponding hydrophobicity scale was defined as $1-(\text{mean RSA})$. The ASA normalization for this calculation used either the empirical or the theoretical scale, evaluated for the ALLOWED region.

Fraction 100\% buried was calculated for each amino acid as the percent of times the program DSSP reported $\text{ASA}<1$\AA\ for each occurrence of that amino acid in the data set. Fraction 95\% buried was calculated for each amino acid as the percent of times that amino acid had an RSA value $<0.05$, where RSA was calculated using the theoretical normalization values of Table~\ref{tab:results} (ALLOWED region).

\subsection*{Data and code availability}
All results and all computer code used to generate these results have been deposited to GitHub.com (https://github.com/mtien/RSA-normalization-values). This inlcudes maximum observed ASA values (both empirical and theoretical) as a function of backbone dihedral angles.

\section*{Acknowledgments}
We thank Jeff Gray for insightful discussions on this work.


\newpage

\section*{Figure captions}

\cleardoublepage

\section*{Tables}

\begin{table}[H]
\caption{\label{tab:results}Proposed values for ASA normalization (in \AA$^2$), compared to previously used scales defined by Rose \emph{et al.}~\cite{Rose1985} and Miller \emph{et al.}~\cite{Miller1987}. Both the theoretical and the empirical scale were evaluated for the ALLOWED region. Corresponding scales evaluated for other regions are provided in Table~\ref{tab:max_SA_complete}.}

\begin{center}
\begin{tabular}{lcccc}
Residue & Theoretical & Empirical & Miller \emph{et al.} (1987) & Rose \emph{et al.} (1985) \\
\hline Alanine & 129.0 & 121.0 & 113.0 & 118.1 \\
Arginine        & 274.0 & 265.0 & 241.0 & 256.0 \\
Asparagine    & 195.0 & 187.0 & 158.0 & 165.5 \\
Aspartate      & 193.0 & 187.0 & 151.0 & 158.7 \\
Cysteine        & 167.0 & 148.0 & 140.0 & 146.1 \\
Glutamate     & 223.0 & 214.0 & 183.0 & 186.2 \\
Glutamine      & 225.0 & 214.0 & 189.0 & 193.2  \\
Glycine          & 104.0 & 97.0 & 85.0 & 88.1  \\
Histidine        & 224.0 & 216.0 & 194.0 & 202.5  \\
Isoleucine     & 197.0 & 195.0 & 182.0 & 181.0  \\
Leucine         & 201.0 & 191.0 & 180.0 & 193.1  \\
Lysine           & 236.0 & 230.0 & 211.0 & 225.8  \\
Methionine   & 224.0 & 203.0 & 204.0 & 203.4  \\
Phenylalanine  & 240.0 & 228.0 & 218.0 & 222.8  \\
Proline & 159.0 & 154.0& 143.0 & 146.8  \\
Serine          & 155.0 & 143.0 & 122.0 & 129.8  \\
Threonine          & 172.0 & 163.0 & 146.0 & 152.5  \\
Tryptophan          & 285.0 & 264.0 & 259.0 & 266.3  \\
Tyrosine          & 263.0 & 255.0 & 229.0 & 236.8 \\
Valine          & 174.0 & 165.0 & 160.0 & 164.5  \\
\hline
\end{tabular}
\end{center}
\end{table}

\begin{table}[H]
\caption{\label{tab:Corr}Absolute value of correlation coefficients $r$ between empirically derived and experimentally derived hydrophobicity scales. The largest significant correlation in each row is highlighted in bold.}
\begin{center}
\footnotesize
\begin{tabular}{lccccc}
 & \multicolumn{5}{c}{Empirical scale} \\ \cline{2-6}\\[-2ex]
Experimental scale & Mean RSA (Rose)$^\text{a}$ & Mean RSA (theor)$^\text{b}$ & Mean RSA (emp)$^\text{c}$ & 100\% buried$^\text{d}$ & 95\% buried$^\text{e}$\\
\hline
Wolfenden \emph{et al.}$^\text{f}$ & 0.614 & 0.681 & 0.681 & \textbf{0.827} & 0.774 \\
Kyte \& Doolittle$^\text{g}$ & 0.841 & 0.879 & 0.881 & \textbf{0.953} & 0.948 \\
Radzicka \& Wolfenden$^\text{h}$ & 0.852 & 0.855 & 0.851 & 0.844 & \textbf{0.888} \\
Moon \& Fleming$^\text{i}$ & 0.704 & 0.748 & 0.752 & 0.678 & \bf{0.764} \\
Fauchere \& Pliska$^\text{l}$ & 0.904 & 0.906 & \textbf{0.910} & 0.734 & 0.878 \\
Wimley \emph{et al.}$^\text{m}$ & 0.463$^\dagger$ & 0.464 & \textbf{0.473} & 0.323$^\dagger$ & 0.417$^\dagger$ \\
MacCallum \emph{et al.}$^\text{k}$ & 0.27$^\dagger$ & 0.265$^\dagger$ & 0.285$^\dagger$ & 0.116$^\dagger$ & 0.227$^\dagger$ \\
\hline
\end{tabular}
\end{center}
{\footnotesize
\begin{raggedright}
$^\text{a}$Mean RSA of residues in protein structures, as calculated by Rose \emph{et al.}~\cite{Rose1985}.\\
$^\text{b}$Mean RSA of residues in protein structures, as given in column 2 of Table~\ref{tab:hydro_scales}.\\ 
$^\text{c}$Mean RSA of residues in protein structures, as given in column 3 of Table~\ref{tab:hydro_scales}.\\
$^\text{d}$Fraction of 100\% buried residues, as given in column 4 of Table~\ref{tab:hydro_scales}.\\
$^\text{e}$Fraction of 95\% buried residues, as given in column 5 of Table~\ref{tab:hydro_scales}.\\
$^\text{f}$Transfer energy from vapor to water~\cite{Wolfenden1981}.\\
$^\text{g}$Hybrid scale based on transfer energy from vapor to water and on the percentages of 95\% and 100\% buried residues in protein structures~\cite{Kyte1981}.\\
$^\text{h}$Transfer energy from cyclohexane to water~\cite{Radzicka1988}.\\
$^\text{i}$$\Delta\Delta G$ between the folded and unfolded state of a mutated membrane-inserted protein, outer membrane phospholipase~A~\cite{Moon2011}. \\
$^\text{k}$Transfer energy calculated from molecular-dynamic simulations of side-chain analogs within a bilayer~\cite{MacCallum2007}.\\
$^\text{l}$Transfer energy between octanol and water~\cite{Fauchere1983}.\\
$^\text{m}$Transfer energy of pentapeptides between octanol and water~\cite{Wimley1996}.\\
$^\dagger$Correlation not statistically significant; all other correlations are significant at $\alpha=0.05$.\\
\end{raggedright}
}
\end{table}

\end{document}